\documentclass[twocolumn,longbibliography,prl,tightenlines,nobalancelastpage,superscriptaddress]{revtex4-2}

\usepackage{amsmath,amssymb}
\usepackage[utf8]{inputenc}
\usepackage[T1]{fontenc}				
\usepackage{lmodern}
\usepackage{fixcmex} 
\usepackage{microtype}
\usepackage{graphicx}
\usepackage[usenames,dvipsnames]{xcolor}
\usepackage{tikz}
\usepackage{bm}
\usepackage[italicdiff]{physics}
\usepackage[per-mode=fraction]{siunitx}
\usepackage{multirow}

\usepackage{capt-of}
\usepackage{dcolumn}
\usepackage{xr}
\definecolor{darkblue}{rgb}{0,0,0.6}
\definecolor{darkred}{rgb}{0.6,0,0}
\definecolor{darkgreen}{rgb}{0,0.6,0}
\usepackage[colorlinks=true,urlcolor=darkblue,citecolor=darkblue,linkcolor=darkred,hyperindex=true,hyperfootnotes=false]{hyperref}

\usepackage[margin=0.75in]{geometry}
\newcommand\thefont{\expandafter\string\the\font}

\newcommand{\bz}{\mathbf{z}}

\makeatletter
\DeclareFontFamily{OMX}{MnSymbolE}{}
\DeclareSymbolFont{MnLargeSymbols}{OMX}{MnSymbolE}{m}{n}
\SetSymbolFont{MnLargeSymbols}{bold}{OMX}{MnSymbolE}{b}{n}
\DeclareFontShape{OMX}{MnSymbolE}{m}{n}{
    <-6>  MnSymbolE5
   <6-7>  MnSymbolE6
   <7-8>  MnSymbolE7
   <8-9>  MnSymbolE8
   <9-10> MnSymbolE9
  <10-12> MnSymbolE10
  <12->   MnSymbolE12
}{}
\DeclareFontShape{OMX}{MnSymbolE}{b}{n}{
    <-6>  MnSymbolE-Bold5
   <6-7>  MnSymbolE-Bold6
   <7-8>  MnSymbolE-Bold7
   <8-9>  MnSymbolE-Bold8
   <9-10> MnSymbolE-Bold9
  <10-12> MnSymbolE-Bold10
  <12->   MnSymbolE-Bold12
}{}

\let\llangle\@undefined
\let\rrangle\@undefined
\DeclareMathDelimiter{\llangle}{\mathopen}%
                     {MnLargeSymbols}{'164}{MnLargeSymbols}{'164}
\DeclareMathDelimiter{\rrangle}{\mathclose}%
                     {MnLargeSymbols}{'171}{MnLargeSymbols}{'171}
\DeclareMathOperator*{\argmin}{arg\,min}
\makeatother

\begin{document}
\title{Natural Selection in the Wake of Catastrophe}
\author{Jesse Young Lin}
\altaffiliation{These authors contributed equally.}
\affiliation{Department of Physics, University of Chicago, Chicago, IL}
\author{Omer Granek}
\altaffiliation{These authors contributed equally.}
\affiliation{Leinweber Institute for Theoretical Physics, University of Chicago, Chicago, IL}
\author{Joshua Sodicoff}
\affiliation{Committee on Genetics, Genomics and Systems Biology, University of Chicago, Chicago, IL}
\author{Seppe Kuehn}
\affiliation{Department of Ecology and Evolution, University of Chicago, Chicago, IL}
\author{David Pincus}
\affiliation{Department of Molecular Genetics and Cell Biology, University of Chicago, Chicago, IL}
\affiliation{Institute for Biophysical Dynamics, University of Chicago, Chicago, IL}
\author{Vincenzo Vitelli}
\affiliation{Department of Physics, University of Chicago, Chicago, IL}
\affiliation{Leinweber Institute for Theoretical Physics, University of Chicago, Chicago, IL}
\affiliation{James Franck Institute, University of Chicago, Chicago, IL}
\date{\today}
\begin{abstract}
Living organisms, from bacteria to humans, are more likely to survive if their traits enhance fitness. In populations well adapted to their environmental niches, natural selection proceeds via rarely beneficial mutations.
But when a
catastrophe 
wipes out niche diversity, sudden adaptation often follows. 
 Here, we present a data-validated 
theory of natural selection in the wake of catastrophe 
 and unveil a simple law that emerges during 
 recovery: 
 the mean fitness 
 relaxes inversely with time, with a prefactor proportional to the number of traits  coupled to the post-catastrophe environment. We put our approach to test using experimental fitness landscapes measured following antibiotic administration to \emph{E. coli}. 
The resulting mean trait adaptation is not described by gradient ascent on a fitness landscape, instead it follows  an algorithm known as Levenberg--Marquardt optimization. Near fitness peaks, evolutionary trajectories are biased against greediness --- from an optimization perspective, post-catastrophic selection is optimistic.
\end{abstract}
\maketitle

\begin{figure*}
\centering
\includegraphics[width=\linewidth]{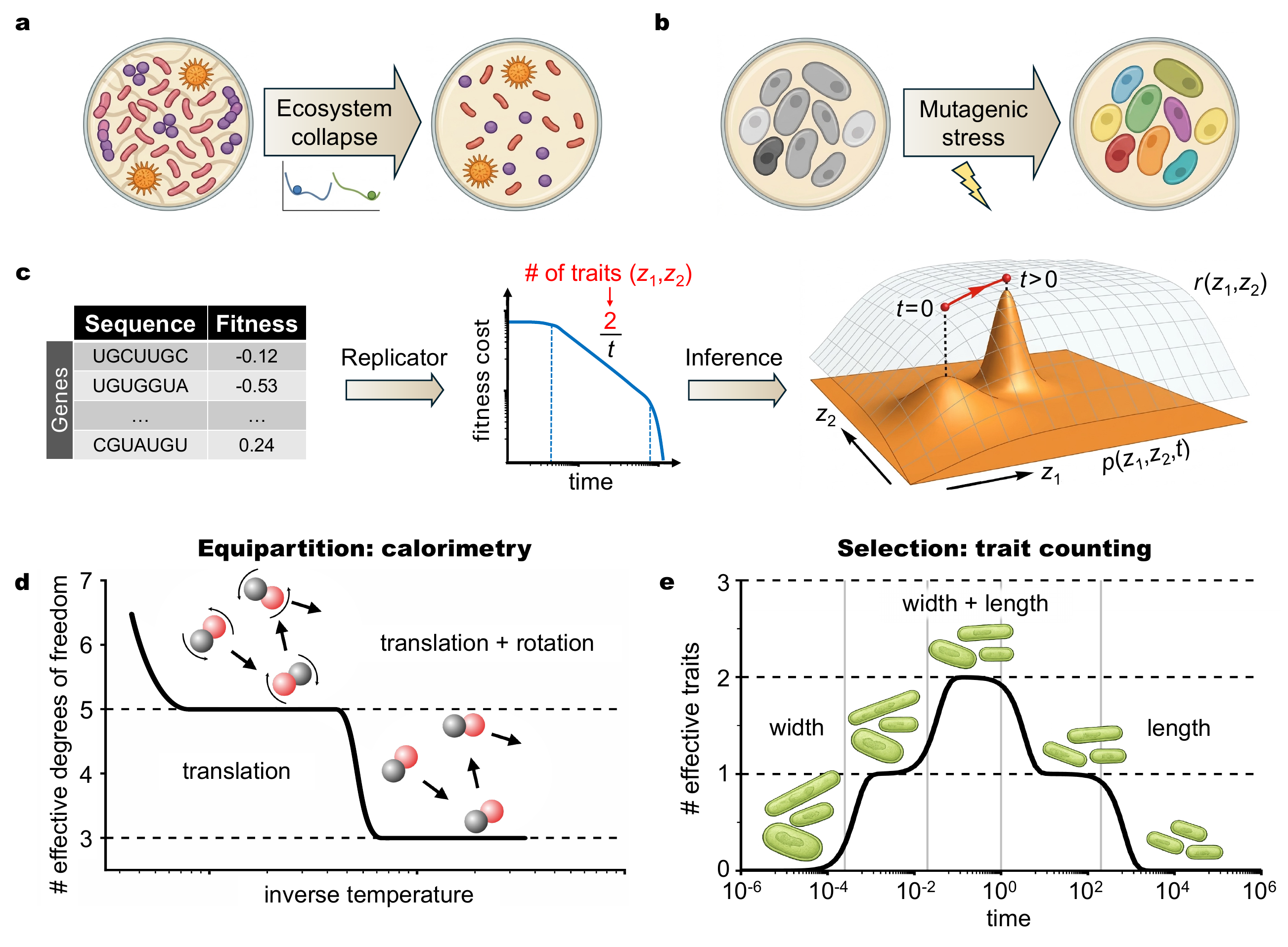}







\centering
\caption{\textbf{Catastrophes turn standing variation into a dynamical probe of fitness-landscape geometry.} \textbf{(a)} Ecosystem collapse after acute stress crosses a tipping point, \textbf{(b)} mutagenic stress causing uncontrolled mutagenesis, and 
\textbf{(c)} Inferring phenotypic fitness landscape geometry from sequencing data through selection on standing variation time traces. A power law regime in the mean fitness cost decay reveals the number of fitness-relevant traits. Initial recovery is dominated by selection on the fitness landscape defined by the trait-dependent rate $r(z_1,z_2)$. Replicator dynamics, Eq.~\eqref{eq:exp}, shifts and contracts the frequency distribution $p(z_1,z_2,t)$ over time. Here, $r(z_1,z_2)=(1+e^{-z_1})^{-1}(1+e^{-z_2})^{-1}$ and $p(z_1,z_2,0)$ is Gaussian. \textbf{(d)} Calorimetry: the specific heat of a diatomic gas plateaus at values encoding the number of thermally excited degrees of freedom, obeying the law of equipartition~\cite{richtmyerIntroductionModernPhysics1947}. \textbf{(e)} Trait counting: the dynamics of the fitness cost for selection on standing variation, Eq.~\eqref{eq:spdef}, predicts time-dependent plateaus encoding the number of relevant phenotypic dimensions under selection. Data are from Eq.~\eqref{eq:freq} with $r(z_1,z_2)=0.1z_1+0.0005z_2$, $N_0(z_1,z_2)=1$ and sampling $z_i=1,2,\ldots,N_i$; $N_1=8000$, $N_2=20000$, with $z_1,z_2$ represented by width and height, respectively.}
    \label{fig:cartoons}
\end{figure*}

Recovery from catastrophe is a collective effort that shapes the evolution of complex systems from  human societies to ecosystems. After an 
ecosystem collapse (Fig.~\ref{fig:cartoons}(a)), a mutagenic stress (Fig.~\ref{fig:cartoons}(b)), an
acute clinical event~\cite{Cox1984AnalysisData,Andersen1993StatisticalProcesses}, 
or a natural hazard~\cite{Narteau2002TemporalRate}, the cohort of affected individuals, be it bacteria or humans, begins its recovery at time $t=0$. What follows is rarely governed by a single, typical timescale: the aggregate response reflects a heterogeneous mix of units with latent ``traits'' (i.e. covariates) $\mathbf{z}=(z_1,\ldots,z_n)$, each failing or rebounding at its own characteristic rate $r(\mathbf{z})$~\cite{Vaupel1979TheMortality,Aalen1988HeterogeneityAnalysis,Hougaard1995FrailtyData,Missov2011AdmissibleAsymptotics,Balan2020AModels}.
For instance, following a layoff, an individual's characteristic time to re-employment $1/r(\bz)$ depends on traits such as individual-specific wage offers or unemployment-benefit duration~\cite{Holmlund1998UnemploymentPractice,Kiefer,Rogerson2005Search-TheoreticSurvey}, while other unmeasured traits are subsumed into a random ``frailty''~\cite{Kiefer,Lancaster1990TheData}. 
The recovery itself is selective: traits with high $r(\bz)$ come to dominate the aggregate response~\cite{Vaupel1979TheMortality,Vaupel1985HeterogeneitysDynamics}. Yet in many settings, the origin of this trait heterogeneity is difficult to disentangle from the complexity of the underlying dynamics~\cite{Kiefer,Lancaster1990TheData,Hougaard1995FrailtyData,Balan2019NonproportionalDifference}.
Here, we ask: what is the average recovery trajectory in the aftermath of catastrophe? 

The answer to this question can be elegantly cast in the language of geometry through the notion of a fitness landscape (see Fig.~\ref{fig:cartoons}(c), rightmost panel). Concretely, a fitness landscape is the surface obtained by assigning to each trait vector $\bz$ a height equal to its rate $r(\bz)$, i.e., the graph of the map $\bz \mapsto r(\bz)$. Recovery corresponds to an optimization flow on this landscape: because higher-$r$ traits recover faster, the trait distribution is progressively reweighted toward higher regions where fitness is greater. As a result, the observed population trajectory reflects both the landscape geometry and the initial trait distribution at $t=0$.



Identifying experimentally relevant traits $\bz$ that define a fitness landscape $r(\bz)$ and the corresponding optimization dynamics calls for data-driven approaches applied to
controlled experimental settings. 
In Fig.~\ref{fig:cartoons}(c), we present an approach that infers the trait dimensionality of the fitness landscape (rightmost panel) starting from experimental fitness data (leftmost panel). Our key insight is that the two are connected by the recovery trajectory of the fitness $r(\bz)$, which we show to exhibit a universal power-law relaxation.
Using a mathematical mapping to the equipartition law of thermodynamics (Fig.~\ref{fig:cartoons}(d)), we show that the recovery provides a quantitative readout of the effective number of traits that determine the fitness landscape (Fig.~\ref{fig:cartoons}(e)), very much like a calorimetry measurement tells us the number of effective degrees of freedom $n$ (e.g. translational, rotational etc.) of a gas whose molar heat capacity is $n R/2$. Once the dimensionality of the fitness landscape is determined, we prove that the resulting mean trait adaptation post-catastrophe is not described by gradient ascent, instead it follows an algorithm known as Levenberg--Marquardt optimization.

Large asexual populations provide a natural testbed for our approach. Their dynamics are normally shaped by slow evolution through the gradual accumulation of mutations~\cite{Good2018EffectiveEvolution,Wortel2023TowardsChallenges,Ascensao2026ExperimentalManipulation}, but their fate is often decided when the environment changes abruptly and survival becomes a race to recover. Such environmental catastrophes can leave behind many competing variants through various routes. 
One route is ecological: a diverse community can be damaged by acute stress~\cite{Hernandez2021EnvironmentalNetworks,Larsen2021OnStudies,Larsen2024Perspective:Transplantation,Hong2024Short-termNetworks,Patnaik2025PredictingReduction,Xu2025Pre-exposureDisturbance}, which may weaken the ecological interactions~\cite{Hildebrand2019Antibiotics-inducedOrder,Thakur2021ResilienceEvent} or precipitate ecosystem collapse~\cite{Dai2012GenericCollapse,Dakos2019EcosystemWorld} (Fig.~\ref{fig:cartoons}(a)).
A second route is genetic: mutagenic stressors, such as ionizing radiation or genotoxic chemicals, can edit the genetic code and rapidly diversify a previously uniform population~\cite{Alcantara-Diaz2004DivergentExposures,Javed2010StrainFermentation,Shibai2017MutationColi} (Fig.~\ref{fig:cartoons}(b)).


\section{Selection on standing variation}
In the regime of \emph{selection on standing variation} (SSV), strong selection acts on many pre-existing phenotypic variants, while \emph{de novo} variation, e.g., due to mutations, is negligible. Building on this simple insight, we develop a general theory of the evolutionary process that emerges in the wake of catastrophe and compare its predictions with existing experimental measurements in genetic landscapes of antibiotic resistance.

The selection-dominated dynamics of the SSV regime are described by the replicator equation~\cite{Nowak2006EvolutionaryDynamics},
\begin{align}
    \partial_t p(\bz,t)=[r(\bz)-\langle r(t)\rangle]p(\bz,t)\;.\label{eq:exp}
\end{align}
Here, $p(\bz,t)$ is the average frequency of the trait (phenotype)
$\bz$ at time $t$ 
and $\langle
O(t)\rangle\equiv\sum_{\bz} p(\bz,t)O(\bz)$ denotes the population mean of an observable~\footnote{Without loss of generality, we consider $n$ non-neutral phenotypic axes in the post-catastrophe environment.}. Equation~\eqref{eq:exp} can describe a host of recovery processes, where the meaning of $r(\bz)$ can range from the Malthusian growth rate under cell division to the stochastic rate for a particular cell-state transition (see Supplementary Information Sec.~9).
It describes pure selection dynamics with vanishing mutation, no ecological interactions, and negligible genetic drift. In Supplementary Information Secs.~1--4, we show that the predictions obtained from Eq.~\eqref{eq:exp} are not limited to pure selection: they persist under weak genetic drift~\cite{7ede0f74-7d87-3888-8232-23a8a77f1a19,Bhat2025APopulations}, weak frequency-dependent selection~\cite{Nowak2006EvolutionaryDynamics}, and weak mutation~\cite{Kimura1965ACharacters,Tsimring1996RNAModel,Gerrish2007CompleteSelection,Sniegowski2010BeneficialPopulations,Desai2011ThePopulations,Alfaro2017Replicator-mutatorFitness,Hamel2020DynamicsLandscape,Mahdisoltani2024MinimalPathogens}. We also delineate the SSV regime quantitatively, identifying the crossovers at which drift, frequency dependence, or mutation cease to be perturbative, and organize the transition to neighboring evolutionary regimes.

The solution to Eq.~\eqref{eq:exp} is given by
\begin{align}
    p(\bz,t)=\frac{N_0(\bz)e^{tr(\bz)}}{Z(t)}\;,\quad Z(t)=\sum_{\bz}N_0(\bz)e^{tr(\bz)}\;,\label{eq:freq}
\end{align}
where $N_0(\bz)$ denotes the population's initial abundances at the onset of
catastrophe.
In what follows, we
elucidate the general behavior of mean observables such as the mean growth rate
$\langle r(t)\rangle$ and the mean trait $\langle\bz(t)\rangle$. 

\section{Scale-free recovery}
As time progresses, if $r^* \equiv r(\bz^*)$ is the dominant accessible fitness peak, then $p(\bz,t)$ increasingly concentrates around the corresponding optimum $\bz^*$ where the fitness cost $D(\bz)\equiv r^*-r(\bz)$ is small. If phenotypic variability
exists near $\bz^*$, the population finely samples an effectively continuous landscape. In this continuum approximation, Eq.~\eqref{eq:freq} becomes
\begin{align}
    Z(t)\sim\int d^n\bz N_0(\bz)e^{tr(\bz)}\sim e^{t r^*}\int dD\,\Omega(D) e^{-tD}\;,\label{eq:cont}
\end{align}
where we introduce the \emph{density of traits},
\begin{align}
    \!\!\!\!\Omega(D) \equiv \sum_{\bz}
N_0(\bz)\delta_{D(\bz), D} \sim\int d^n\bz N_0(\bz)\delta[D(\bz)-D].\!\label{eq:DOT}
\end{align}

The asymptotic relaxation towards optimality can be characterized
systematically by applying Laplace's method to Eq.~\eqref{eq:cont} in the limit $t\rightarrow\infty$ (see Methods). Near optimality, where $\delta z=\abs{\bz - \bz^*}$ is small, the fitness cost is given by $D \sim \delta z^Q$, and $\Omega \sim D^{n/Q-1}$ due to $\dd[n]{\bz} \propto D^{n/Q - 1} \dd{D}$. Laplace's method and the identity $\langle r(t)\rangle=\frac{d}{dt}\log Z(t)$ then provide the
universal asymptotic form (see Methods),
\begin{align}
\ev{D(t)} &= r^*-\langle r(t)\rangle
          = \frac{n}{Q t} + \order{t^{-2}}\;.\label{eq:plaw}
\end{align}
Equation~\eqref{eq:plaw} is the first main result: post-catastrophic recovery is slow and scale-free, with a universal, inverse-time decay of the mean fitness cost. This contrasts with the exponential relaxation observed when a beneficial trait fixes in a system with low standing variation. The prefactor depends solely on the local shape parameter $n/Q$, where $n$ is the number of non-neutral phenotypic axes near the optimum and $Q$ characterizes the local curvature of the peak~\cite{Tenaillon2014TheGenetics}. For multiple accessible optima, the expansion holds locally:
well-separated peaks produce piecewise algebraic relaxation, punctuated by
switches between the optima, with an appropriate local value of
$n/Q$ at each stage (Supplementary Information Sec.~5). 

The universality of Eq.~\eqref{eq:plaw} admits a simple analogy with equilibrium statistical
mechanics. The distribution $p(\bz,t)\propto N_0(\bz)e^{-tD(\bz)}$ is analogous to the
equilibrium distribution of a coarse-grained variable $\bz$ at temperature $T=1/t$ in the
energy landscape $D(\bz)$. In this language, Equation~\eqref{eq:plaw} is the equipartition law:
to leading order, each degree of freedom contributes $T/Q$ to the mean energy~\cite{huangStatisticalMechanics1987}.
Calorimetry relies on equipartition to probe the energy-landscape geometry~\cite{richtmyerIntroductionModernPhysics1947}: intermediate
plateaus in specific heat curves probe the effective number of degrees of freedom
(Fig.~\ref{fig:cartoons}(d)). By the same
logic, a catastrophe may act as a dynamical probe of the dimensionality of the underlying fitness landscape: $n/Q$ can be
measured directly through a kinetic fitness assay at multiple time points, agnostic to the underlying phenotype space. In this experiment, the \emph{specific fitness},
\begin{align}
c(t)\equiv \frac{d}{d(1/t)}\langle D(t)\rangle
= t^2\left[\langle D^2(t)\rangle-\langle D(t)\rangle^2\right],\label{eq:spdef}
\end{align}
plateaus at $n/Q$ in the scale-free regime (Fig.~\ref{fig:cartoons}(e)). As in gas
calorimetry, where plateaus of the specific heat are finite intervals between
mode-activation crossovers, the algebraic decay of Eq.~\eqref{eq:plaw} and the
plateau of $c(t)$ are observed in
an intermediate asymptotic regime $\tau_{\rm s}\ll t\ll \tau_{\rm
c}$~\cite{barenblattScalingSelfsimilarityIntermediate1996}.
Here, $\tau_{\rm s}$, defined by $\ddot{c}(\tau_{\rm s})\equiv0$, marks the end of the
initial transient, while $\tau_{\rm c}\sim1/\Delta r$ marks the breakdown of the continuum
approximation, beyond which selection occurs between a small, discrete number of traits
separated by a characteristic fitness gap $\Delta r$ (see Methods for precise
form). Independent trait subspaces with additive fitness effects may acquire
distinct intermediate asymptotic plateaus, giving rise to the ``pyramid'' in
Fig.~\ref{fig:cartoons}(e). In Supplementary Information Sec.~6 we show using a scaling theory that this
phenomenology extends to weakly-correlated trait subspaces.

%

In the intermediate asymptotic regime, knowing one of $n$ or $Q$ determines the
other via the measurement of $n/Q$.
Generically, $Q=1$ is expected for directional selection, where $\bz^*$ lies at the boundary of the sampled
trait space, whereas $Q=2$ is expected for stabilizing selection, where $\bz^*$ lies in its interior~\cite{Lande1983TheCharacters}.
Technical exceptions to this rule are discussed in Methods. Fixing $Q$ according
to these guidelines and biological insight, the phenotypic dimension $n$, also known as the phenotypic complexity, can be
inferred purely from fitness data.

Moreover, the independent growth in the idealized SSV regime implies that Eq.~\eqref{eq:freq} holds at all levels of description: from genotype through mesoscopic phenotypes to complex macroscopic phenotypes. We thus follow the approach in Figure~\ref{fig:cartoons}(c) to probe the phenotypic dimension in genotypic fitness assays using SSV. We apply this framework to two large experimental datasets in \textit{E.
coli}~\cite{Otto2024AEnvironment,
Papkou2023ALandscape2} which allows us to identify candidates for the relevant phenotypic axes near optimality.

\section{Phenotypic dimension inference}
\begin{figure*}
    \centering
    \includegraphics{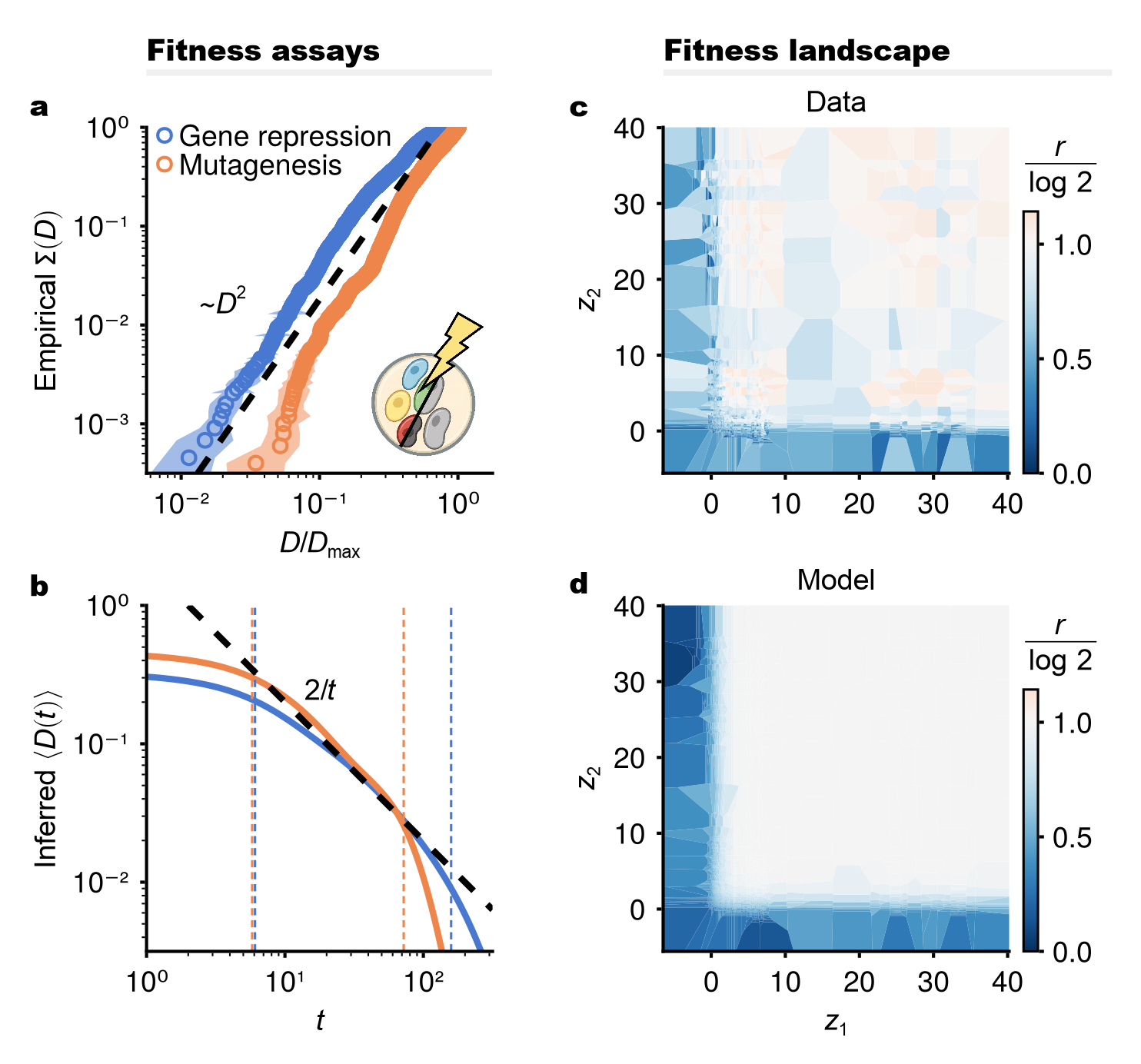}
    \caption{\textbf{Static fitness assays in \emph{E. coli} measure the phenotypic dimension relevant to evolutionary response.} 
   Two complementary assays measure $n/Q$, where $n$ is the number of independent
traits under selection and $Q$ describes the curvature of the fitness peak. Static
assays screen a whole library of variants at once, giving a distribution of
fitness costs relative to the best variant; the cumulative fraction of variants
within a rescaled fitness cost $\rho = D/D_\textrm{max}$ is predicted to scale as $\Sigma(\rho) \sim 
\rho^{n/Q}$. 
Kinetic assays track a recovering population over time, giving the mean
remaining fitness cost $\ev{D(t)}$ which is predicted to decay as $(n/Q)/t$ (Eq.~\eqref{eq:plaw}; the $1/t$ slope is universal). 
The two are interconvertible by a Laplace transform (Eq.~\eqref{eq:cont}). 
\textbf{(a, b)} Antibiotic fitness landscapes measured statically by Otto et al. \cite{Otto2024AEnvironment} via
gene-repression ($\mathcal{N} = 4464$) and Papkou et al. \cite{Papkou2023ALandscape2} \emph{folA} mutagenesis ($\mathcal{N} = 261332$);
\textbf{(a)} the measured distribution (open circles), \textbf{(b)} the same data mapped to
kinetic form (solid lines). Dashed lines are theory with
no fitting parameters; vertical dashed lines mark $\tau_\textrm{s}$ and $\tau_\textrm{c}$, between which the law applies. Both are consistent with $n/Q = 2$, i.e. two-trait landscapes
with monotonic peaks.
\textbf{(c,d)} Gene repression assay data and modeling~\cite{Otto2024AEnvironment} indicate a two-dimensional and monotonic fitness landscape $r(z_1,z_2) = (1 + e^{-z_1})^{-1}(1+e^{-z_2})^{-1}$.
    }
    \label{fig:data}
\end{figure*}



Figure~\ref{fig:data} juxtaposes two static assays~\cite{Otto2024AEnvironment,
Papkou2023ALandscape2} of antibiotic fitness landscapes in \emph{E. coli}. Static
assays measure fitness values across an ensemble of types, yielding $\Omega(D)$.
These data determine the kinetic assay, which tracks a recovering population over time, measuring $Z(t)$ or equivalently
$\ev{r(t)}=\frac{d}{dt}\log Z(t)$. The two assays are linked by the Laplace
transform, Eq.~\eqref{eq:cont}, analogous to the equivalence of the
microcanonical and canonical ensembles in statistical mechanics. For both datasets, the static and kinetic assays are consistent
with Eq.~\eqref{eq:plaw} and with the common shape parameter $n/Q=2$. We now
show that in each case the local landscape is consistent with $Q=1$ and $n=2$.

Figure~\ref{fig:data}(a-b) reanalyzes static fitness assays under antibiotic stress. The first is a multi-gene double repression assay~\cite{Otto2024AEnvironment}. Using the CRISPR-i method, the authors imposed variable repression levels $R_i,R_j\in[0,1]$ on genes $i,j$ from a pool of nine genes, where $R_i=0$ denotes unperturbed expression and $R_i=1$ complete knockdown. The experiment was found to be well-described by the 
fitness model,
\begin{align}
    r_{ij}(R_{ij},R_{ji})=\frac{1}{1+e^{\kappa_i(R_{ij}^{\rm eff}-R_i^0)}}\frac{1}{1+e^{\kappa_j(R_{ji}^{\rm eff}-R_j^0)}},\label{eq:otto}
\end{align}
where $R_{ij}^{\rm eff}=a(R_j)R_i$ are effective repression levels renormalized
by gene-gene interactions $a(R_j)$. We therefore identify the fitness landscape
$r(z_1,z_2)=(1+e^{-z_1})^{-1}(1+e^{-z_2})^{-1}$ as a function of
the effective expression phenotypes $z_1(i,R_{ij}^{\rm eff})=\kappa_i
(R^0_i-R^{\rm eff}_{ij})$ and $z_2(j,R_{ji}^{\rm eff})=\kappa_j (R^0_j-R^{\rm
eff}_{ji})$ (see Fig.~\ref{fig:data}(c-d)). 

The monotonicity of the landscape implies $Q=1$, while the two coordinates imply
$n=2$, consistent with Fig.~\ref{fig:data}(c-d). Furthermore, specific genetic
backgrounds correspond to interpretable subspaces of the $(z_1,z_2)$ landscape.
By conditioning on repression of gene $i$, we obtain a reduced landscape for
which we infer $n/Q$ using the standard Clauset--Shalizi--Newman method (CSN), which provides a maximum likelihood estimate over
bootstrap-resampled landscapes~\cite{clausetPowerlawDistributionsEmpirical2009}
(see Methods). Then, genetic backgrounds with $n/Q$ lower than the wild type
correspond to phenotypic subspaces with potentially reduced dimensions. For
example, while the full repression dataset yields $n/Q=1.9\pm0.28$, conditioning
on repression of \emph{gdhA} or \emph{gltB} yields $n/Q=1.48\pm0.27$ and
$1.76\pm0.45$, respectively. These scalar readouts reflect broad, bimodal
bootstrap distributions, arising from the experimental sampling of two distinct
regions of the effective landscape $r(z_1,z_2)$: an effectively one-dimensional
region, where a neutral direction is present, and a two-dimensional region,
where it is absent (Supplementary Information Sec.~7). Repeating the analysis of
Fig.~\ref{fig:data} in these backgrounds reveals that the
\emph{gdhA}/\emph{gltB}-conditioned landscapes are predominantly one-dimensional
(Extended Figs.~5-6). This is consistent with the known
\emph{gdhA}/\emph{gltB} interaction: repression of either gene alone has little
fitness effect, whereas only combined repression is
lethal~\cite{Cote2016TheColi2,Otto2024AEnvironment}. Thus, conditioning on either
\emph{gdhA} or \emph{gltB} probes a region of the effective landscape
$r(z_1,z_2)$ with a neutral direction.

The second assay is a combinatorially-complete, pooled fitness assay of an \emph{E. coli} \emph{folA} gene segment under trimethoprim antibiotic stress~\cite{Papkou2023ALandscape2}. Using CRISPR-Cas9, the authors generated an \emph{in vivo} library spanning $99.7\%$ of the genotypic variants at nine nucleotide positions encoding amino acid sites $26$--$28$ of dihydrofolate reductase (DHFR), a central metabolic enzyme targeted by trimethoprim. Figure~\ref{fig:data}(a-b) shows this landscape is consistent with $n/Q=2$. We argue $Q=1$, and hence $n=2$, on kinetic grounds as follows. Enzymatic activity $v$, whose contribution to growth is mediated by metabolic flux~\cite{Kacser1981THEDOMINANCE,Dean1986FitnessColi,Dykhuizen1990EnzymeSolution,Newton2018EnzymeComplicated}, is captured, in the absence of inhibitor,  by the prototypical Michaelis-Menten model
\begin{align}
    E + S \;\underset{k_{-1}}{\overset{k_1}{\rightleftharpoons}}\; ES \;\overset{k_{\rm cat}}{\rightarrow}\; E + P\;.
\qquad
v=\frac{k_\text{cat}[E][S]}{K_{\rm M}+[S]}\;.
\end{align}
Here, $[E]$ and $[S]$ are the intracellular enzyme and substrate concentrations and $K_{\rm M} = (k_{-1}+k_{\mathrm{cat}})/k_1$ is the Michaelis constant. Under a competitive inhibitor $I$, such as trimethoprim, $K_{\rm M}$ is replaced by $K_{\rm M}^{\rm  app}=K_{\rm M}(1+[I]/K_{\rm i})$, where $K_{\rm i}$ is the inhibition constant. The monotonicity of $v\!=\!v(k_{\rm cat}, \!K_{\rm M},\!K_{\rm i},\![E],\![S],\![I])$ implies $Q=1$ near optimality. The readout $n/Q=2$ in Fig.~\ref{fig:data} therefore implies $n=2$ relevant trait axes. We thus hypothesize the phenotypic landscape $v(z_1,z_2)=z_1/(1+z_2)$, where $z_1=k_\text{cat}[E]$ and $z_2=K_{\rm M}^{\rm  app}/[S]$ are plausibly distinct axes. This is supported by large-scale enzyme surveys, which find only weak global correlation between $k_{\rm cat}$ and $K_{\rm M}$~\cite{Bar-Even2011TheParameters}, and DHFR mutational studies, which show that nearby active-site substitutions can shift kinetic parameters in various directions~\cite{Tamer2019High-OrderSelection,Krucinska2022Structure-guidedPathogens}. 
This dimensional reduction recurs in kinetic modeling of DHFR mutations~\cite{Rodrigues2016BiophysicalResistance,Nguyen2024TheInteraction}, minimal models of drug resistance~\cite{Das2020PredictableTradeoffs2,Das2022DrivenEnvironments,Das2025Epistasis-mediatedTradeoffs}, and  pharmacodynamic descriptions~\cite{Regoes2004PharmacodynamicRegimens,Das2020PredictableTradeoffs2}. More broadly, data-driven models suggest that $n=2$ collective coordinates can capture functional variation across protein families~\cite{Ziegler2023LatentSpace,Gaszek2025Higher-orderResistance2}.

Microscopically, the phenotypic dimension $n=2$ is also supported by the enzyme structure~\cite{davidStructureFunctionAlternative1992} and the topology of the fitness basins~\cite{Papkou2023ALandscape2}. The presence of aspartic acid and glutamic acid at site 27 (D27/E27) is one of the primary distinctions between bacterial and vertebrate DHFRs, respectively, yet the ligand-bound wildtype and D27E structures are nearly identical: although glutamic acid includes an additional methylene group in its side chain, it extends away from the binding site~\cite{davidStructureFunctionAlternative1992}. Consistent with this, Ref.~\cite{Papkou2023ALandscape2} reports that the fitness basin for DHFR variants with either D27 or E27 is smooth and continuous. This is supported by our CSN analysis, which finds the phenotypic dimension conditioned on D27 or E27 to be $n/Q=1.92 \pm 0.15,\;1.82 \pm 0.49$ respectively, while combining the two conditioned spaces increases it to $n/Q=2.57 \pm 0.23$. The increase reflects reduced constraint in the shared smooth basin and the reduced tendency of mutations at sites $26$ and $28$ to dramatically affect function.

By contrast, cysteine at site 27 (D27C) leads to a different active site architecture: the conformation of the backbone remains similar, but the cysteinyl sulfur atom accepts a hydrogen bond from T113, bridging the active site and constraining the $\alpha$-helix and $\beta$-sheet holding C27 and T113, respectively~\cite{davidStructureFunctionAlternative1992}. The thiol group of C27 is critical for substrate protonation, substituting for the function of the carboxyl group present in aspartic and glutamic acid~\cite{davidStructureFunctionAlternative1992}. This binding-site constraint is consistent with selection analysis in Ref.~\cite{Papkou2023ALandscape2} and our reanalysis: D27C variants form a rougher, separate basin with $n/Q=1.67 \pm 0.31$, and site 28 is enriched for leucine after selection relative to D27E or wildtype backgrounds. When conditioned on the presence of L28, the C27 basin increases in average fitness and measured dimension, with $n/Q=1.86\pm0.20$. Thus, before site 28 fixes, the dominant phenotypic axis is site 28 itself, whereas after fixation, the smoother basin resembles the D/E27 basin. While this dimensionality readout does not identify molecular phenotypes explicitly, it reveals their conditional structure and provides insight into their microscopic origins. 

\begin{figure*}[!t]
    \centering
    \includegraphics{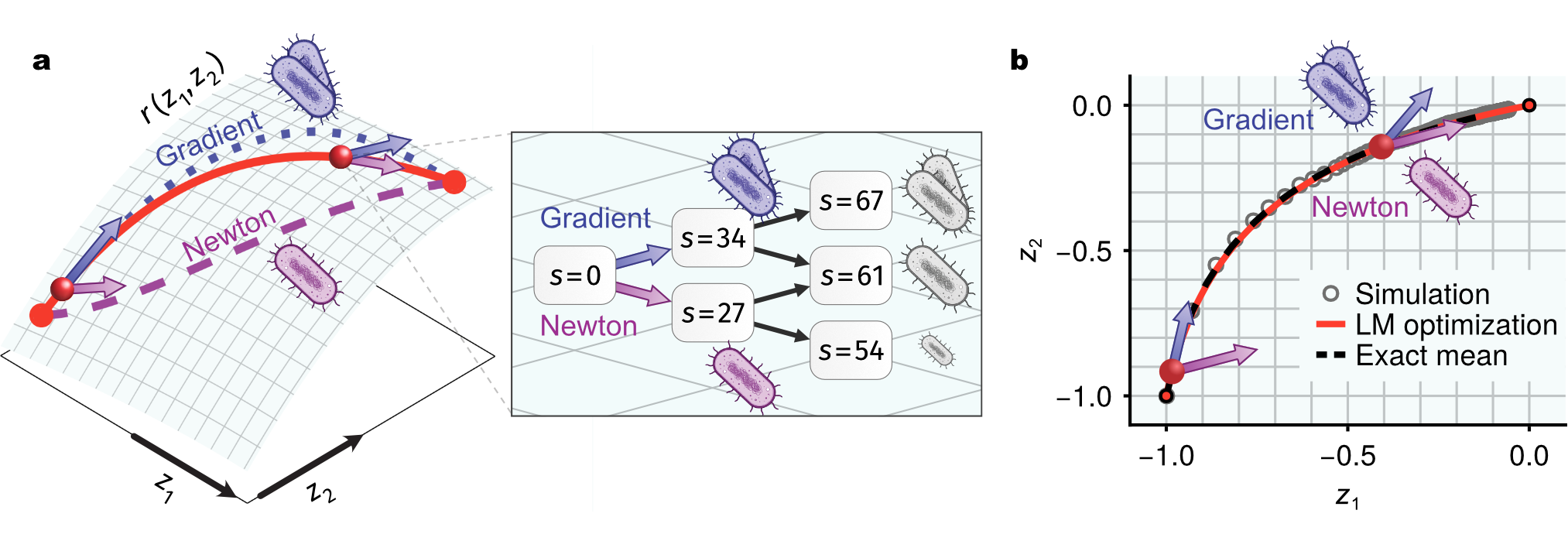}
    \caption{\textbf{Selection on standing variation follows curvature-sensitive trajectories on fitness landscapes.} \textbf{(a)} Gradient ascent, Newton's method, and Levenberg-Marquardt dynamics on the landscape $r(z_1,z_2)=10-z_1^2-4z_2^2+0.1(z_1^2+z_2^2)^2$, and the Levenberg-Marquardt dynamics which interpolates between them.
    The Levenberg-Marquardt dynamics adaptively selects between the steepest ascent (gradient, blue) and a curvature-sensitive ascent (Newton, purple), with directions shown by arrows. While gradient flow is preferable on the steep fitness slopes, Newton's method is preferable near the peak where some directions may be shallow. Newton's method
    reweights the diminishing fitness gradient and takes shorter paths in anisotropic landscapes. Inset: Discretization of the trajectory about $\bz_0 = (-0.42,-0.14)$ as a selection ``tournament'' (lattice spacing $a=5\times10^{-3}$). The winner of each round is determined by the optimization rule and the selection coefficient $s(\bz) = [r(\bz)-r(\bz_0)]/\kappa_\textrm{max} a^2$ where $\kappa_\textrm{max}$ is the largest eigenvalue of $-H_{ij}(\bz^*)$ (rounded to the nearest integer for visualization). The gradient rule (blue) chooses the high-$s$ successor, whereas the Newton rule (purple) considers the next generation and may locally prefer the less fit successor.
    \textbf{(b)} Highest-frequency trait trajectories $\bz(t)$ on the
    fitness landscape in (a). Stochastic birth-death simulations (circles) are compared with the Levenberg-Marquardt dynamics (Eq.~\eqref{eq:LM}) and the exact solution
    (Eq.~\eqref{eq:freq}; lines). For sufficiently
    large and peaked initial populations, $\ev{\bz(t)}$ converges to the Levenberg-Marquardt
    trajectory.}
    \label{fig:LM}
\end{figure*}

The above analyses go beyond previous attempts to measure the phenotypic dimension using effective phenotypic evolution models~\cite{LeNagard2013SelectionBasedNetworks,Tenaillon2014TheGenetics,Schneemann2024FishersSpeciation}. The SSV regime makes phenotypic dimension measurable from the idealized recovery dynamics alone, agnostic to the unknown genotype--phenotype and phenotype--fitness maps required in mutation-limited settings~\cite{Blanquart2014PropertiesModel,Tenaillon2014TheGenetics,Blanquart2016EpistasisModel,Hwang2017GenotypicModel,Ascensao2026ExperimentalManipulation}.
Moreover, it enables a direct comparison between information content and phenotypic dimension. The information content of a library of $\mathcal{N}$ variants is $L\equiv\log_2\mathcal{N}$~\cite{Schneider2000EvolutionInformation}. The landscapes of Refs.~\cite{Otto2024AEnvironment,Papkou2023ALandscape2} analyzed in Fig.~\ref{fig:data}(a-b) have information content $L\simeq12,18$, respectively. A non-redundant sampling of a phenotypic space of dimension $n$, characteristic size $X$, and mesh size $a$ gives $\mathcal{N}\sim(X/a)^n$, or equivalently $L\sim \log(X/a)n$. Our analysis yields $L/n=6,9$ for the respective landscapes, consistent with limited (subexponential) redundancy of traits in the sampled space. This is also consistent with the phenotypic map $(i,j,R_i,R_j)\mapsto z_k$ found for Ref.~\cite{Otto2024AEnvironment}, which displays strong phenotypic epistasis, whereby the effect of a mutation depends strongly on its background, and pleiotropy, whereby a mutation can affect several traits simultaneously~\cite{Stearns2010OneRetrospective,Tenaillon2014TheGenetics}. Despite the lack of direct access to the genotype--phenotype map, our measurement of the phenotypic dimension for Ref.~\cite{Papkou2023ALandscape2} likewise supports the presence of these effects.

\section{Non-gradient optimization}
The inverse-time law in Eq.~\eqref{eq:plaw} constrains the scalar recovery. A complementary question is therefore how the dominant trait itself evolves in phenotype space. Let $\bz=\bz(t)$ denote the highest-frequency trait, around which $p(\bz,t)$ concentrates. In the SSV regime this trait climbs toward higher fitness, but not generally by steepest ascent (Fig.~\ref{fig:LM}(a)). Maximizing the distribution in Eq.~\eqref{eq:freq} implies
\begin{align} 
\partial_i [S(\bz)+t  r(\bz)] = 0\;,\label{eq:sp_timedep} 
\end{align} 
where $S(\bz) \equiv \log
N_0(\bz)$. Differentiating Eq.~\eqref{eq:sp_timedep} with respect to time yields
\begin{align}
-\qty(\partial_i \partial_j S(\bz) + t \partial_i \partial_j r(\bz))\dv{z_j}{t} = \partial_i r(\bz)\;,\label{eq:LM}
\end{align}
where summation over repeated indices is implied. 
The solution $\bz(t)$ to Eq.~\eqref{eq:LM} recovers the long-time asymptotics of Eq.~\eqref{eq:plaw}. Moreover, $\ev{\bz(t)}\sim\bz(t)$ when $N_0(\bz)$ is sufficiently peaked around an initial trait $\bz_0$, as verified  by the stochastic simulations in Fig.~\ref{fig:LM}(b)
and analytically in the Supplementary Information Sec.~8.

Equation~\eqref{eq:LM} resembles a gradient ascent modified with a
time-dependent friction tensor $M_{ij}(\bz,t)= -\partial_i \partial_j
[S(\bz) + t  r(\bz)].$ It can be recast as a function of $s=\log t$ as
\begin{align}
-\qty(e^{-s}\partial_i \partial_j S(\bz) + \partial_i \partial_j r(\bz))\dv{z_j}{s} = \partial_i r(\bz)\;,\label{eq:LMs}
\end{align}
which can be identified as the continuous-time limit of the
Levenberg-Marquardt algorithm, a regularized version of Newton's
method~\cite{Levenberg1944ASquares,Marquardt1963AnParameters,Attouch2011AInclusions,Transtrum2012ImprovementsMinimization2,nesterovLecturesConvexOptimization2018}. 
As demonstrated in Fig.~\ref{fig:LM}(a) for $n=2$, the Levenberg-Marquardt
trajectory initially follows steepest ascent. As time proceeds, however, the motion becomes increasingly sensitive to landscape curvature through the fitness Hessian $H_{ij}\equiv \partial_i\partial_j r(\bz)$, and approaches Newton's method near the optimum. For $n=1$, this affects only trajectory speed; for $n>1$, it changes the path itself.

To illustrate this, we discretize Eq.~\eqref{eq:LM} on a square
lattice of spacing $a$. The fitness
gradient on the right-hand side of Eq.~\eqref{eq:LM} becomes $\partial_i r(\bz) \to r(\bz + \vb{e}_i) - r(\bz)$, while the
fitness Hessian becomes $H_{ij}(\bz)\to r(\bz+\vb e_i+\vb e_j)-r(\bz+\vb e_i)-r(\bz+\vb e_j)+r(\bz)$, where $(\vb e_i)_j=a\delta_{ij}$. Figure~\ref{fig:LM}(a) then recasts the discrete-time update as a selection tournament.
Under the gradient rule, the preferred successor $z_i$ is the one with the larger pairwise fitness gain $\partial_i r$. Under the Newton rule, the preferred successor depends on three-point comparisons through $-(H^{-1})_{ij}
\partial_j r$, and can therefore locally favor the less fit successor. Strikingly, this remains the case
even if the fitter trait has fitter successors. 

Predicting post-catastrophic trajectories thus requires more than pairwise fitness differences: three-point comparisons are needed to evaluate local curvature.
The full Levenberg-Marquardt trajectory interpolates smoothly between these two schemes and takes progressively greater fitness risks.
Near the fitness peak, selection therefore becomes biased against greediness, preferentially exploring locally shallower phenotypic directions---a property known in optimization theory as ``optimism''. Thus, from an optimization perspective, post-catastrophic selection on standing variation is optimistic~\cite{bertsekasReinforcementLearningOptimal2019}.

\section{Acknowledgements}
\begin{acknowledgements}
This research was partly supported by the National Science Foundation through
the Physics Frontier Center for Living Systems (PHY2317138) as well as NSF (DMS-2235451) and Simons
Foundation (MPS-NITMB-00005320) to the NSF-Simons
National Institute for Theory and Mathematics in Biology (NITMB). V. V. is a Chan Zuckerberg
Biohub Chicago Investigator. O.G. acknowledges support from
the Leinweber Institute for Theoretical Physics, the Center for Living Systems at The University of Chicago and a MRSEC-funded Kadanoff–Rice
fellowship from The University of Chicago Materials Research
Science and Engineering Center, which is funded by NSF
(DMR-2011854).
\end{acknowledgements}

\onecolumngrid
\vspace{5mm}

\section*{Methods}

\newcommand{\showsection}{1} 

\ifnum\showsection=1

\subsection{Laplace's method}
Laplace’s method is a systematic asymptotic expansion of Laplace integrals about the dominant rate.
For the partition function $Z(t)$, inserting $\Omega(D) = \Omega_0 D^{n/Q - 1} + \order{D^{n/Q}}$ into Eq.~\eqref{eq:cont} yields
\begin{align}
Z(t) &\sim e^{tr^{*}}\int \dd{D}\, \Omega(D)e^{-tD} 
 =e^{tr^{*}}\Omega_{0}\Gamma(n/Q)t^{-n/Q}\qty[1+\order{t^{-1}}]\;,
\label{eq:Laplace integral}
\end{align}
which provides Eq.~\eqref{eq:plaw} after taking the logarithmic derivative. The geometric interpretation of Laplace's method as a locally smooth approximation of the fitness landscape is more apparent by
directly expanding the multidimensional integral in Eq.~\eqref{eq:cont}. With $\delta z=|\delta\bz|$ and $\delta\bz\equiv\bz-\bz^*$, the leading-order fitness cost can be expanded as
\begin{align}
    D(\delta \bz) = \delta z^Q D_0\left(\frac{\delta \bz} {\delta z}\right) +\delta z^{Q+1} D_1\left(\frac{\delta \bz} {\delta z}\right)+ \ldots\;,\label{eq:rhomogeneous}
\end{align}
where $Q>0$. Substituting Eq.~\eqref{eq:rhomogeneous} into Eq.~\eqref{eq:cont} and applying the change of variables $\delta \bz=t^{-1/Q}\bm{\xi}$, we obtain for the regular, smooth cases described below,
\begin{align}
Z(t) = e^{tr^{*}}N_{0}({\bz}^{*})\qty(\int \dd[n]{\bm{\xi}}e^{-\xi^{Q}D_0\left(\frac{\bm{\xi}}{|\bm{\xi}|}\right)}) t^{-n/Q} \left[1+\mathcal{O}(t^{-1})\right]\;,\label{eq:Zexp}
\end{align}
where the integration over $\bm{\xi}$ is constrained to the sampled sector of the phenotype space relative to $\bz^*$. Comparing with Eq.~\eqref{eq:Laplace integral} determines $\Omega_0 = \frac{N_0(\bz^*)}{\Gamma(n/Q)}\int \dd[n]{\bm{\xi}}e^{-\xi^{Q}D_0\left(\frac{\bm{\xi}}{|\bm{\xi}|}\right)}.$ The $\mathcal{O}(t^{-1})$ correction in Eq.~\eqref{eq:Zexp} is obtained for smooth landscapes with a generic interior maximum ($Q=2$) or boundary vertex maximum ($Q=1$). It is likewise obtained for a maximum in the interior of an $m$-dimensional boundary. In that case, one may choose $n-m$ coordinates normal to the boundary and $m$ coordinates tangent to it. Then, the density of traits has $m$ quadratic contributions and $n-m$ linear contributions, again reproducing the $\mathcal{O}(t^{-1})$ correction. In the most general case of a nonsmooth landscape, the correction can remain $\mathcal{O}(t^{-1/Q})$, leading to a $\mathcal{O}(t^{-1-1/Q})$ correction in Eq.~\eqref{eq:plaw}. Smooth but degenerate maxima with $Q>2$ provide another technical exception, with fractional subleading powers even when the landscape is differentiable.

An analogous calculation shows that the population average $\ev{O(t)}$ of an arbitrary smooth observable $O(\bz)$ in an arbitrary landscape is given by 

\begin{align}
\ev{O(t)} = O(\bz^*) +\begin{cases} 
t^{-1/Q} \langle \bm{\xi}\rangle_Q \cdot \bm{\nabla} O(\bz)|_{\bz=\bz^*}+\mathcal{O}(t^{-2/Q})\;,&\langle\bm{\xi}\rangle_Q\neq0\\
t^{-2/Q} \left[ \langle \xi_i\xi_j\rangle_Q\left(\frac12 O_{ij} + O_i S_j\right) -\langle \xi_i\xi^{Q+1} D_1(\bm{\xi}/\xi)\rangle_Q O_i  \right]+\mathcal{O}(t^{-3/Q})\;,&\langle\bm{\xi}\rangle_Q=0
\end{cases}\;,
\label{eq:Laplace integral obs}
\end{align}
where $O_i\equiv\partial_iO|_{\bz=\bz^*}$, $O_{ij}\equiv\partial_i\partial_jO|_{\bz=\bz^*}$, $S_i\equiv\partial_i\log N_0(\bz)|_{\bz=\bz^*}$, $\xi=|\bm{\xi}|$, and
\begin{align}
\langle f(\bm{\xi})\rangle_Q \equiv\frac{\int d^n\bm{\xi}e^{-\xi^Q D_0\qty(\frac{\bm \xi}{\xi})}f(\bm{\xi})}{\int d^n \bm{\xi}e^{-\xi^Q D_0\qty(\frac{\bm \xi}{\xi})}}\;.
\end{align}
The expansions in Eq.~\eqref{eq:Laplace integral} and Eq.~\eqref{eq:Laplace
integral obs} indicate that the power law regime generically depends on $N_0(\bz)$ and the observable $O(\bz)$. Indeed, in the most general case, the integral in Eq.~\eqref{eq:Laplace integral obs} is sensitive to the local geometry about $\bz^*$. Like the correction in Eq.~\eqref{eq:Zexp}, the leading decay in Eq.~\eqref{eq:Laplace integral obs} generically becomes $\mathcal{O}(t^{-1})$ in the regular cases, both in boundary ($Q=1$, $\langle \bm{\xi}\rangle_1\neq0$) and interior ($Q=2$, $\langle \bm{\xi}\rangle_2=0$) cases. Fractional powers arise in nonregular exceptions, including nonsmooth landscapes and smooth but degenerate maxima.

\subsection{Trait counting and intermediate asymptotics}
The early $\tau_{\textrm s}$ and late $\tau_{\textrm c}$ timescales defining the intermediate asymptotic regime $\tau_{\rm s}\ll t\ll \tau_{\rm c}$
correspond to the initial and final inflection points of the specific fitness $c(t)$, between which a $c
\sim n/Q$ plateau is obtained.

Fluctuations in real data may cause numerous inflection points in $c(t)$, despite a single equipartition plateau. The first inflection point $\tau_{\rm s}$, can be found by solving $\ddot{c}(\tau_{\textrm s}) = 0$ numerically.  For a characteristic fitness gap \(\Delta r\)
near the optimum $r=r^*$, the final inflection
point $\tau_{\rm c}\sim 1/\Delta r$ can be determined via the late-time expansion \(Z(t)/e^{tr^*} = 1 + e^{- t\Delta r} +
\order{e^{ -2\Delta rt}}\).

A more accurate determination of $\tau_{\rm c}$ employs a criterion from the theory of Bose-Einstein
condensates~\cite{holthausUniversalRenormalizationSaddlepoint1999,polychronakosRoleDensityStates2026}.
Splitting the partition function into discrete and continuous parts
\begin{align}
\frac{Z(t)}{e^{tr^*}} = \sum_{\delta r = 0}^{\delta r_{\textrm c}} e^{-t \delta r}
+ \int_{\delta r_{\textrm c}}^\infty \dd{\delta r}\Omega(\delta r)e^{-t \delta
r}
\end{align}
The time $\tau_{\textrm c}$ is then the crossover point where the
discrete and continuous distributions balance,
\begin{align}
\sum_{\delta r = 0}^{\delta r_{\textrm c}} e^{-\tau_{\textrm c} \delta r} =
\int_{\delta r_{\textrm c}}^\infty \dd{\delta r}\Omega(\delta
r)e^{-\tau_{\textrm c} \delta r}\;. \label{eq:BE timescale}
\end{align}
By Laplace's method, the right-hand side satisfies $\int_{\delta r_{\textrm c}}^\infty
\dd{\delta r}\Omega(\delta r)e^{-\tau_{\rm c} \delta r} \sim \tau_{\rm c}^{-1}\Omega_0
\delta r_{\textrm c}^{n/Q-1}e^{-\tau_{\rm c} \delta r_{\textrm c}}$ and vanishes if $\tau_\textrm{c}
\to \infty$. In contrast, the left-hand side approaches unity. Therefore, $\delta r_{\textrm
c}$ must be small enough that $\int_{\delta r_{\textrm c}}^\infty
\dd{\delta r}\Omega(\delta r) > \sum_{\delta r = 0}^{\delta r_{\textrm c}} 1$
which guarantees that Eq.~\eqref{eq:BE timescale} has a solution. In
Fig.~\ref{fig:data}(b) we solve Eq.~\eqref{eq:BE timescale} numerically with  $\delta r_{\textrm c} = \Delta r$.

The above construction explains how several plateaus can appear in a single readout of $c(t)$. Consider trait subspaces $\bz=(\bz_1,\ldots,\bz_m)$ whose fitness effects are additive over the relevant range, $r(\bz)=\sum_\alpha r_\alpha(\bz_\alpha)$, and whose initial distributions are factorizable, $p_0(\bz)=\prod_\alpha p_{0,\alpha}(\bz_\alpha)$. Then, Eq.~\eqref{eq:spdef} for subspace $\alpha$, implies
\begin{align}
c(t)=\sum_\alpha c_\alpha(t)\;,
\end{align}
where we denote the contribution by subspace $\alpha$ using a subscript. Each subspace thus has its own intermediate asymptotic regime $\tau_{{\rm s},\alpha}\ll t\ll \tau_{{\rm c},\alpha}$ in which $c_\alpha(t)\simeq n_\alpha/Q_\alpha$. The specific fitness measures the sum of the contributions under SSV at that time (Fig.~\ref{fig:cartoons}(e)), much akin to calorimetry, where different mechanical modes are thermally activated over different temperature regimes (Fig.~\ref{fig:cartoons}(d)). In Supplementary Information Sec.~6, we show that the above analysis extends to weakly correlated subspaces with non-additive effects using a scaling theory.

\subsection{Dimension estimation by Clauset--Shalizi--Newman power-law fitting}
Given experimental measurements $\{r_i\}_{i=1}^\mathcal{N}$ of a fitness landscape over types $i$, we define the fitness cost
$D_i \equiv r^*-r_i$, where $r^*\equiv \max_j r_j$. In the scale-free regime, the integrated density of traits satisfies
\begin{align}
    \Sigma(D) \equiv \sum_i \Theta(D-D_i)\sim D^{n/Q}\;,
\end{align}
thus providing a readout of
the shape parameter $n/Q$ by the predicted linear relationship $\log \Sigma(D)
\sim \frac{n}{Q} \log D$.

To provide a principled error estimate for this
readout, we use the CSN power-law maximum-likelihood procedure~\cite{clausetPowerlawDistributionsEmpirical2009,gillespieFittingHeavyTailed2015}. If the power law $\Sigma(D)\sim D^{n/Q}$ holds up
to a cutoff scale $D_\textrm{c}$, the maximum likelihood estimator is
\begin{align}
\widehat{n/Q} = \mathcal{N}_{\rm c}\left[\sum_{0<D_i\leq D_\textrm{c}}\log\!\left(\frac{D_{\rm c}}{D_i}\right)\right]^{-1}\;,
\label{eq:CSN_MLE}
\end{align}
where $\mathcal{N}_{\rm c}\equiv\sum_{0<D_i\leq D_{\rm c}}$ and the cutoff scale $D_{\rm c}$ is selected to minimize the Kolmogorov--Smirnov statistic,
\begin{align}
D_\textrm{c} \equiv \argmin_{D_c'} \max_{D < D_\textrm{c}'} \abs{\mathcal{N}_\textrm{c}^{-1}\eval{\Sigma(D)}_{D < D_\textrm{c}'} - \qty(\frac{D}{D_\textrm{c}'})^{n/Q}}\;.
\end{align}
The uncertainty in $D_{\rm c}$ is quantified by bootstrapping with many resamples of the data. Here, we use
bootstrapping with $10^{4}$ resamples and report the bootstrap
average and standard deviation. This gives an
objective, dataset-agnostic statistical procedure to estimate $n/Q$. Because reduced landscapes conditioned on specific genetic backgrounds contain fewer points, their estimates are more sensitive to the inferred cutoff $D_{\rm c}$, which sets the dynamic range of the
data that contributes to the estimate. In these cases, the bootstrap distribution itself is informative; broad or multimodal distributions are inspected separately in Supplementary Information Sec.~7.

\fi


\subsection{Stochastic simulations}
To show that our results deriving from Eq.~\eqref{eq:freq} are robust to
demographic noise, we conduct a stochastic simulation of a population growing
with trait-dependent birth rates $r(\bz)$ given by the fitness landscape of
Fig.~\ref{fig:LM}(a) and vanishing death rates, resulting in the trajectory shown in Fig.~\ref{fig:LM}(b). We
use a standard Gillespie $\tau$-leaping
scheme~\cite{gillespieApproximateAcceleratedStochastic2001} with $\tau = 0.1$
and couple the population to a chemostat by periodically diluting it back to its initial population $Z(0) =
10^6$ every $\Delta t = 1$ time units, congruent with standard experimental setups in microbial
evolution~\cite{lenskiDynamicsAdaptationDiversification1994}. The initial population $N_0(\bz)$ is an isotropic Gaussian
of variance $\sigma^2 = 1/24$, ensuring convergence to Eq.~\eqref{eq:LM} (Supplementary Information Sec.~9). An empirical population drawn from the distribution $N_0(\bz)$ has negligible probability of providing fine sampling near the
optimum. We therefore utilize an importance-biased Monte Carlo
technique~\cite{elgartRareEventsStatistics2004,bucklewIntroductionRareEvent2004}
by simulating from an initially uniform distribution over the traits and
estimating the average trait averaged over trajectories via
\begin{align}
    \overline{\ev{\bz}}
= \frac{\int \dd[n]{\bz} \bz N_0(\bz) \overline N_\text{sim}(\bz,t)}{\int
\dd[n]{\bz} N_0(\bz) \overline N_\text{sim}(\bz,t)}\;,
\label{eq:rare_event_sim}
\end{align}
where $N_\text{sim}(\bz,t)$ is the simulated population. This estimator is asymptotically
exact for large populations, where $p_{\rm sim}(\bz,t)=\overline N_\text{sim}(\bz,t)/\sum_\bz'\overline N_\text{sim}(\bz',t)$ converges to the
replicator solution, Eq.~\eqref{eq:freq}. Further details are given in Supplementary Information Sec.~9.
\vspace{5mm}
\twocolumngrid

\bibliographystyle{apsrev4-2}
\bibliography{references,references_JL,references_extras}

\end{document}